\documentclass[aps,prl,preprint,superscriptaddress]{revtex4-1}
\usepackage{amsmath,amssymb}
\usepackage{bm}
\usepackage{graphicx,color,natbib}
\usepackage{placeins}
\usepackage[dvipsnames]{xcolor}
\usepackage{array}

\newcommand{\beq}{\begin{equation}}
\newcommand{\eeq}{\end{equation}}
\newcommand{\beqa}{\begin{eqnarray}}
\newcommand{\eeqa}{\end{eqnarray}}

\newcommand{\paddyspeaks}[1]{{\color{black} #1}}
\newcommand{\paddycomment}[1]{{\color{black} #1}}
\newcommand{\ioatzinspeaks}[1]{{\color{black} #1}}

\newcommand{\pecl}{\operatorname{\mathit{P\kern-.08em e}}}

\begin{document}

\title{Decorated Protein Networks: Functional Nanomaterials with Tunable Domain Size}

\author{Ioatzin R\'ios de Anda}
\affiliation{H.H. Wills Physics Laboratory, Tyndall Avenue, Bristol, BS8 1TL, UK}

\author{Ang\'elique Coutable-Pennarun}
\affiliation{BrisSynBio Synthethic Biology Research Centre, Life Sciences Building, Tyndall Avenue, Bristol, BS8 1TQ, UK}
\affiliation{School of Cellular and Molecular Medicine, University Walk, Bristol, BS8 1TD, UK}

\author{Chris Brasnett}
\affiliation{H.H. Wills Physics Laboratory, Tyndall Avenue, Bristol, BS8 1TL, UK}

\author{Stephen Whitelam}
\affiliation{Molecular Foundry, Lawrence Berkeley National Laboratory, Berkeley, California 94720, USA}

\author{Annela Seddon}
\affiliation{H.H. Wills Physics Laboratory, Tyndall Avenue, Bristol, BS8 1TL, UK}
\affiliation{Centre for Nanoscience and Quantum Information, Tyndall Avenue, Bristol, BS8 1FD, UK}
\affiliation{\paddyspeaks{Bristol Centre for Functional Nanomaterials, University of Bristol, Bristol, BS8 1TL, UK}}

\author{John Russo}
\affiliation{School of Mathematics, University Walk, Bristol, BS8 1TW, UK}

\author{J.L. Ross Anderson}
\affiliation{School of Biochemistry, University of Bristol, Bristol, BS8 1TD, UK}
\affiliation{School of Cellular and Molecular Medicine, University Walk, Bristol, BS8 1TD, UK}

\author{C. Patrick Royall}
\affiliation{H.H. Wills Physics Laboratory, Tyndall Avenue, Bristol, BS8 1TL, UK}
\affiliation{School of Chemistry, University of Bristol, Cantock's Close, Bristol, BS8 1TS, UK}
\affiliation{Centre for Nanoscience and Quantum Information, Tyndall Avenue, Bristol, BS8 1FD, UK}

\begin{abstract}
The \paddycomment{implementation of natural and artificial } 
proteins with designer properties and functionalities offers unparalleled opportunity for functional nanoarchitectures formed through self-assembly. However, to exploit the opportunities offered we require the ability to control protein assembly into the desired architecture while avoiding denaturation and therefore retaining protein functionality. Here we address this challenge with a model system of fluorescent proteins. Using techniques of self-assembly manipulation inspired by soft matter where interactions between components are controlled to yield the desired structure, we show that it is possible to assemble networks of proteins of one species which we can decorate with another, whose coverage we can tune. Consequently, the interfaces between domains of each component can also be tuned, with applications for example in energy transfer. Our model system of fluorescent proteins eGFP and mCherry retain their fluorescence throughout the assembly process, thus demonstrating that functionality is preserved. 
\end{abstract}

\maketitle

Complex, hierarchical, long-range ordered materials with building blocks at the nano-- and micro--scale can be obtained by controlling self-assembly, through careful manipulation of the interactions between the assembling components. If the constituents of such structures present optical, magnetic, electric, chemical or biological properties, then the assemblies formed hold great potential for a myriad of applications in a great variety of fields, including  photonics, energy transfer and storage, magnetic levitation, catalysis, drug delivery, tissue scaffolding, cancer treatment and gene transfection. \cite{xu2016,bai2016,luo2016,kobayashi2018,yang2016}. Perhaps the greatest source of inspiration for the design and construction of functional materials at the nanoscale is nature itself. The exquisite level of complexity, specificity, efficiency and sophistication of nature's structures makes their building blocks (proteins, polymers, nucleic acids, carbohydrates and lipids) an attractive possibility to be exploited for new devices. Of these building blocks, proteins exhibit the largest diversity of structures and functions and thus have the greatest potential to be exploited as the functional components for novel nanostructured materials \cite{bai2016,luo2016,vo-dinh2005,watkins2017,kobayashi2018,yang2016}. They are capable of carrying out structural, catalytic, transport, packaging, optical, specific recognition, electrical, information storage and metabolic functions \cite{luo2016,bai2016}.

Like colloidal suspensions, protein solutions \paddyspeaks{may} 
be stable or, if the interactions between the proteins are suitably controlled, they exhibit aggregation, gelation and crystallisation \cite{dumetz2008,mcmanus2016,zhang2007,wang2012}. In nature, aggregation is often avoided, so that proteins will form a condensed phase only if 
(\emph{i}) they have evolved to do so (i.e. viral capsids), if 
(\emph{ii}) mutations lead to mis-folding or to interactions that produce aggregation (i.e. Alzheimer's disease), or if 
(\emph{iii}) changes in the medium (i.e. ionic strength or massive changes in temperature or pH) are induced \cite{mcmanus2016,wang2012,saric2014} so that in the latter two scenarios, the proteins lose their functionality.

\paddyspeaks{By driving assembly in a controlled fashion, it is possible to generate architectures at the nanoscale, which may be subsequently functionalised.} An example of the latter structures are gels, where the components (colloids or proteins) are partly associated or interconnected with each other through chemical or physical links, forming ramified amorphous percolating networks \cite{zaccarelli2007,varrato2012,royall2018}. However, unlike most gels found in nature, here we aim to produce new complex and functional gel networks with a binary system in which the domain size of each species can be controlled \cite{whitelam2014}. By incorporating multi-enzyme cascades, light harvesting arrays and electron transfer proteins, such multicomponent networks hold potential as advanced materials for catalysis, energy transduction and nanoscale electronics. 

\begin{figure}
	\centering
	\includegraphics[height=0.18\textheight]{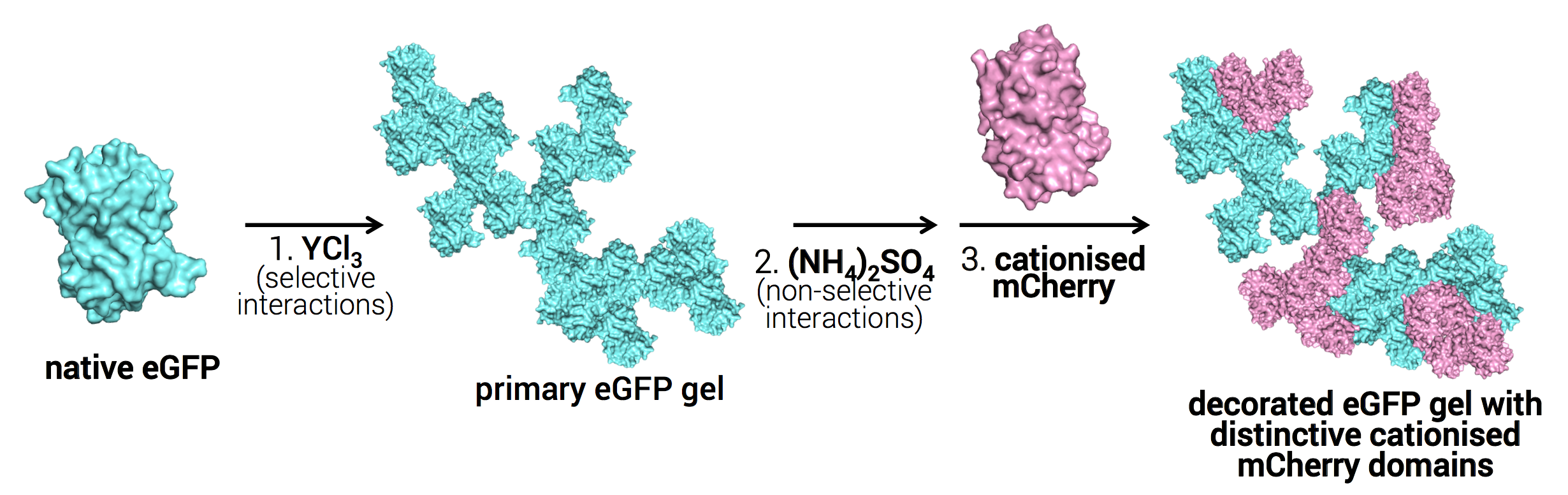}
	\caption{\footnotesize{Strategy to yield decorated \paddyspeaks{protein networks}. 
	Specific interactions of trivalent ions and protein surface modification (cationisation) are exploited to provide and control the specificity of protein-salt interactions to yield two-step decorated gel with distinguishable protein domains. 
	}}
	\label{fig:general}
\end{figure}

While so-called bigels of two distinct unconnected networks have been produced with colloids \cite{dimichele2013} and (denatured) proteins \cite{blumlein2015}, here our focus is on a single network of two components, where we can control the assembly such that the domain size and hence the \emph{interfaces} between domains can be tuned. In particular,  we aim to \paddyspeaks{self-assemble} 
decorated gel networks formed by distinguishable protein domains, where these preserve their native (and functional) structure. The fluorescent proteins enhanced green fluorescent protein (eGFP) and mCherry were chosen as the model system due to the ease of monitoring the assembly process and retention of protein function with optical microscopy.

To \paddyspeaks{assemble} 
these structures, we exploit the effects of salts on protein solutions \cite{elcock2001,zhang2007,duong-ly2014,okur2017,piazza2000} based on specific salt-protein interactions. Trivalent ions have been shown to selectively interact with the surface-exposed carboxylic groups of the acidic residues of the proteins, which in turn, produces gelation or crystallisation \cite{zhang2010,zhang2011,roosen-runge2013,zhang2014b}. Thus, by effectively modifying the surface of the proteins, we provide the specificity required to gain control over the interactions that lead to gelation as shown in Fig. \ref{fig:general}. Once we determined the response of the interactions to control parameters such as salt concentration, we show that control over protein-salt interactions allowed tuning the primary gel coverage with the secondary protein and the sizes of protein domains within the gel structure.

To our knowledge, these are the first structures where both the gel components retain functionality, i.e., retain their native state, and where control over their size within the gel is possible. As such, the strategies proposed herein represent a promising route to yield a new type of functional biomaterial.

\section*{Methods}
The fluorescent proteins eGFP and mCherry were obtained through bacterial expression in E. coli, followed by cell separation, cell lysis and centrifugation. The proteins where further purified using Ni-NTA affinity chromatography. To gain better control over protein-salt interactions, mCherry was cationised through the addition of hexamethylendiamine (HMDA) and 1-Ethyl-3-(3-dimethylaminopropyl)carbodiimide (EDC) to a native protein solution in order to change the negative charge of the native protein to a positive one, without affecting its functionality. We obtained the decorated networks following the methodology shown in Fig. \ref{fig:general}. Briefly, a primary gel of native eGFP is formed by adding yttrium chloride. Then, we add ammonium sulphate to precipitate cationised mCherry. 
Different concentration ratios of native eGFP and cationised mCherry were tested. The structures were immediately characterised using confocal laser scanning microscopy. Finally, image analysis was performed to quantify the size of individual protein domains within the decorated gel.

\section*{Results}

We present our results over three main sections. The first corresponds to increasing the specificity of protein-salt interactions to gain sufficient control over the gelation process. We then proceed to assemble the desired structures following the two-step addition of native and modified proteins to different salts. Finally, we implement our strategy to tune the protein domain sizes within the gel.

\subsection{Increasing Specificity of Protein 
Interactions: Designing the Protein System}

To prevent non-specific interactions between the salts and the proteins, we require high protein-salt specificity. We followed the work of Zhang and coworkers, \paddyspeaks{who} investigated specific interactions between trivalent salts and the acidic residues of globular proteins \cite{roosen-runge2013}. In particular, we tested different concentrations of iron chloride (FeCl$_{3}$) and yttrium chloride (YCl$_{3}$) with eGFP and mCherry. For \paddyspeaks{iron chloride} 
we observed protein denaturation (Fig. \textbf{S1}), likely due to a pH decrease produced by the acidic Fe$^{3+}$ ion \cite{roosen-runge2013}. For YCl$_{3}$, the results are illustrated in Fig. \textbf{S2 a} where eGFP forms gels readily whilst retaining its fluorescence, and thus, its native structure. For mCherry, only clusters of the fluorescent protein were observed (Fig. \textbf{S2 b}). In agreement with previous work on BSA \cite{zhang2010}, re-entrant solution was observed for mCherry at 5 mM YCl$_{3}$. \paddyspeaks{We therefore proceed with YCl$_{3}$ to assemble the eGFP into a network.}

These studies on trivalent cations have also shown that proteins with positively charged surfaces do not gel regardless of the particular salt and concentration used \cite{zhang2010}. By cationising mCherry we prevent its interaction with YCl$_{3}$ and further increase the selective interactions of this salt with our system components. We determined the protein $\zeta$-potential after cationisation and we did observe a charge inversion for the cationised protein ($\zeta_\mathrm{cat-mCherry}$ = +9.3 mV) from its native negative counterpart ($\zeta_\mathrm{mCherry}$ = -7.0 mV).

We continued to test if we have effectively  ``blocked'' the interaction between YCl$_{3}$ and the protein. To do this, we mixed solutions of cationised mCherry against the same concentrations of salts as for the native proteins. We did not observe any gelation for the cationised protein at any concentration tested, as shown in Fig. \textbf{S3 a}. Finally, we \paddyspeaks{investigated whether} 
the addition of positive charges to the protein affected its gelation with ammonium sulphate, (NH$_\mathrm{4}$)$_\mathrm{2}$SO$_\mathrm{4}$. The Hofmeister Series, classifies salts based on their 
to stabilise or de-stabilise protein solutions \cite{okur2017}. Ammonium Sulphate precipitates proteins via non-specific preferential solvation and as such, is widely used for protein purification, precipitation and storage protocols \cite{okur2017,wingfield2001,piazza2000}.  To destabilse cationised mCherry, we added 3 M of Ammonium Sulphate to a 8 mg/mL solution of the protein. Images of the results obtained are shown in Fig. \textbf{S3 b} where the cationised protein formed gel structures without disrupting the protein conformation.

We thus obtained a system where native eGFP forms a gel through specific interactions with YCl$_{3}$, whereas cationised mCherry does not. However, the latter still gels readily with (NH$_\mathrm{4}$)$_\mathrm{2}$SO$_\mathrm{4}$. We used these specific and differential interactions to yield the decorated gels.

\begin{figure*}[h!]
	\centering
	\includegraphics[height=0.27\textheight]{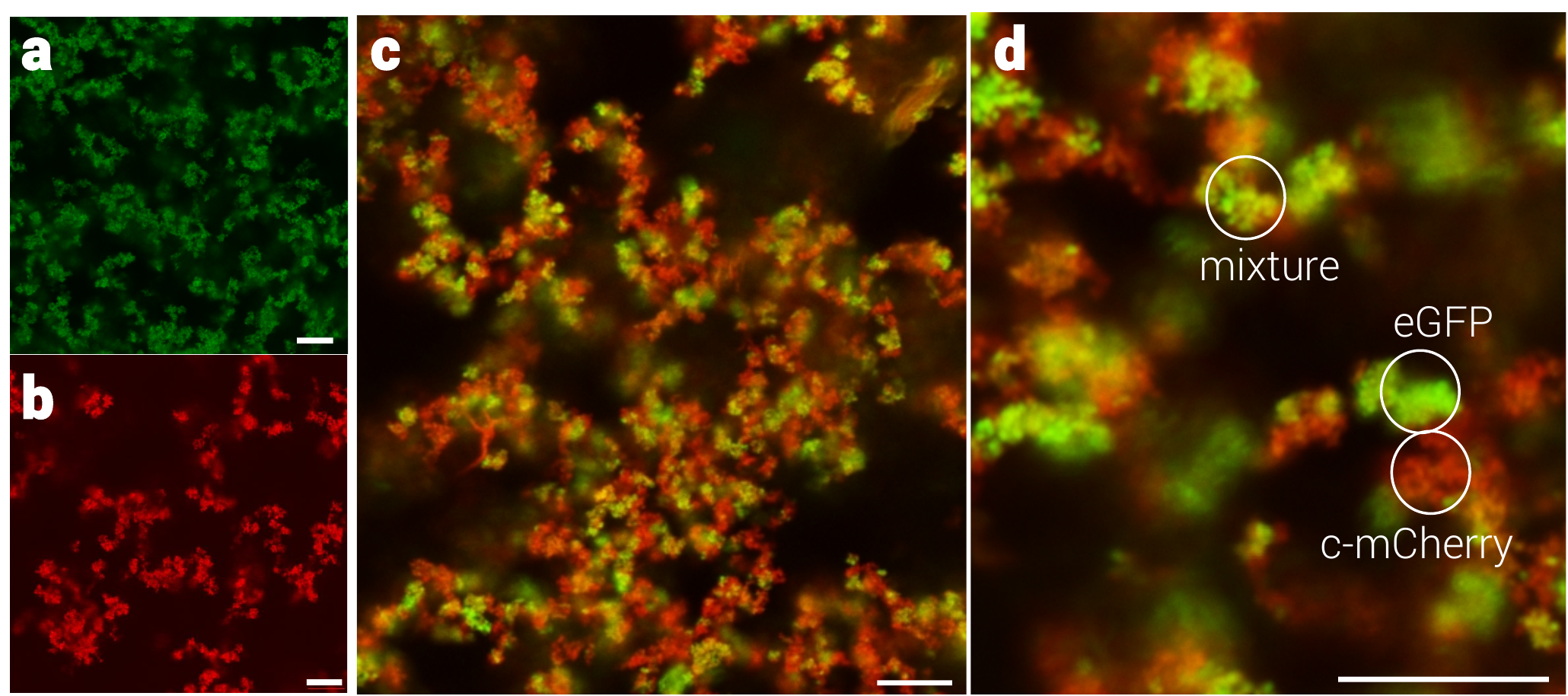}
	\caption{\footnotesize{Confocal images of individual and decorated eGFP and cationised mCherry gels. (a) Native eGFP gel formed through the addition of yttrium chloride. (b) Cationised mCherry gel formed by adding ammonium sulphate. (c-d) Confocal images of the expected decorated gel networks with distinctive native eGFP (green) and cationised mCherry (red) domains. Native eGFP and was mixed on its own with yttrium chloride to form a gel. Ammonium sulphate is added followed by cationised mCherry. Scale bars = 10 $\mu$m. 
	}}
	\label{fig:Prot-Bigel-ScndStrat}
\end{figure*}

\subsection{Two-step formation of decorated protein network}

\paddyspeaks{Previous investigations of} 
protein precipitation have found that the presence of a protein undergoing gelation might trigger co-gelation (\emph{i.e.} co-precipitation) of other oppositely charged proteins in solution, due to complex interactions between the species \cite{konno2001}. Although the surface charge of our proteins is of only a few eC (-1.46 eC for eGFP and +2.71 eC for cationised mCherry), in order to prevent such co-gelation between them, we decided to first form a gel with the native protein and YCl$_{3}$. We then added (NH$_\mathrm{4}$)$_\mathrm{2}$SO$_\mathrm{4}$ to gel cationised mCherry, which we added at last, as illustrated in Fig. \ref{fig:general}.

The results of this strategy are shown in Fig. \ref{fig:Prot-Bigel-ScndStrat} \textbf{a} and \textbf{b}, where 
the desired decorated gel \paddyspeaks{structures} were successfully obtained. In Fig. \ref{fig:Prot-Bigel-ScndStrat}  we can clearly distinguish green and red-only domains (native eGFP and cationised mCherry, respectively), which indicate that they are solely composed of one of the proteins. Thus, the strategy proposed successfully produced decorated gels, with distinctive protein domains where both components have kept their functional structure. However, there are still large yellow areas (overlay of green and red channels), which indicate that cationised mCherry deposited directly on the surface of the preexisting native eGFP gel, areas which we will refer as ``mixture'' in the following sections. Both observations are clearer at higher magnifications (Fig. \ref{fig:Prot-Bigel-ScndStrat} \textbf{b}).

\ioatzinspeaks{We further confirmed that the strategy followed above to increase salt-protein specificity is indeed required to yield the decorated structures. We mixed native eGFP and mCherry solutions and added several (NH$_\mathrm{4}$)$_\mathrm{2}$SO$_\mathrm{4}$ salt concentrations. The results are shown in Fig. \textbf{S5}. No gelation occurs before 1 M of (NH$_\mathrm{4}$)$_\mathrm{2}$SO$_\mathrm{4}$, and we only start observing small clusters formed solely by eGFP up to 1.5 M of (NH$_\mathrm{4}$)$_\mathrm{2}$SO$_\mathrm{4}$. However, we start observing co-gelation of the proteins at 1.6 M (NH$_\mathrm{4}$)$_\mathrm{2}$SO$_\mathrm{4}$, where the proteins are undistinguishably mixed on the lengthscales we access. This is more evident as the salt concentration increases. Full protein co-gelation is observed at 3 M of (NH$_\mathrm{4}$)$_\mathrm{2}$SO$_\mathrm{4}$. Both proteins retained their fluorescence yielding functional binary gels formed by a mixture of both proteins, as they appear yellow, with no domains of individual proteins clearly identified.}

\begin{figure*}
	\centering
	\includegraphics[height=0.65\textwidth]{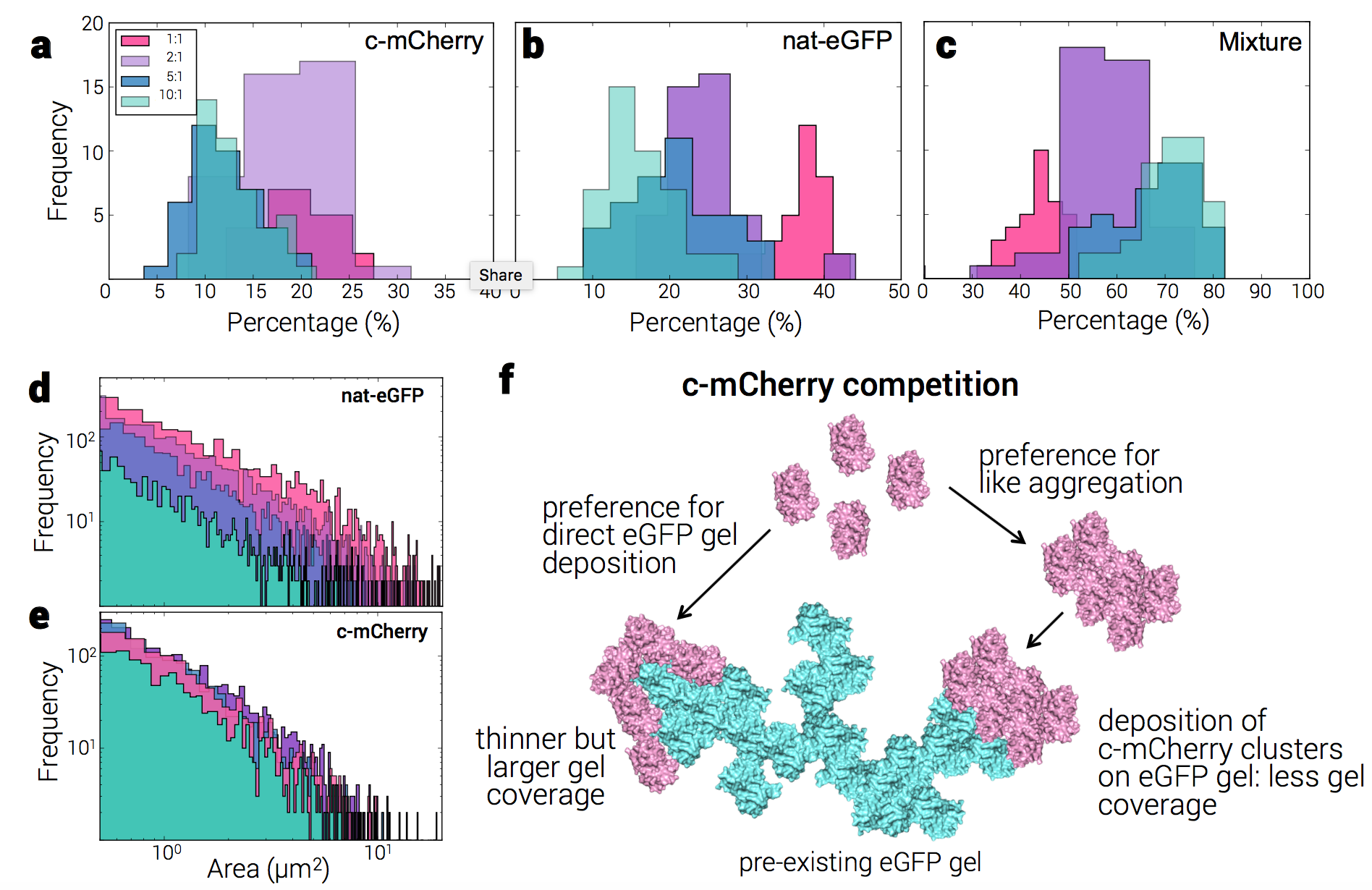}
	\caption{\footnotesize{Comparison of the distribution of the percentages of individual regions of cationised mCherry (a), native eGFP (b) and coated gel (c) in the bi-network when the native eGFP: cationised mCherry concentration ratio of the proteins is varied from 1:1 (pink), 2:1 (purple), 5:1 (blue) and 10:1 (turquoise). 
(d) and (e) Comparison of the distribution of the domain area size of individual regions of native eGFP (b) and cationised mCherry (e) in the decorated gels. The native eGFP: cationised mCherry concentration ratio is varied as in (a). 
(f) Cationised mCherry competition between depositing directly on the pre-existing eGFP gel or forming clusters before their deposition.}}
	\label{fig:all}
\end{figure*}

\subsection{Tuning the coverage of primary eGFP gel with cationised mCherry}

\paddyspeaks{Having assembled the desired architecture,}
we then tested whether we could control the coverage of the native eGFP gel and the domain sizes of the regions with individual native eGFP and cationised mCherry. To do this, we decreased the concentration of cationised mCherry to a half, a fifth and a tenth of the eGFP concentration, which was kept at 4 mg/mL. We hypothesise that changes in these values might provide an indication of which of the interactions is preferred as follows: if the amount of regions identified as cationised mCherry increases or remains the same as the quantity of this protein is reduced, it might suggest that there is a slight preference for intra-protein interactions. On the contrary, if the gel coverage augments as less cationised mCherry is added, then the inter-protein interactions are preferred.

Figure \ref{fig:all} \textbf{a-c} show the results obtained, where Fig. \ref{fig:all} \textbf{a} corresponds to the different percentages according to the different ratios tested for cationised mCherry, Fig. \ref{fig:all} \textbf{b} corresponds to native eGFP, whereas Fig. \ref{fig:all} \textbf{c} corresponds to the gel coverage. The different native eGFP and cationised mCherry ratios 1:1, 2:1, 5:1 and 10:1 are coloured in pink, purple, blue and turquoise, respectively. Unexpectedly, as we decreased the amount of cationised mCherry, so did the percentage of uncovered native eGFP gel (Fig. \ref{fig:all} \textbf{b}), whereas the percentage of covered gel increased (Fig. \ref{fig:all} \textbf{c}). No significant differences are observed between 5:1 and 10:1 ratios of protein concentration (Fig. \ref{fig:all} \textbf{a-c} blue and turquoise colours). Although a similar trend to ``free" native eGFP is observed for cationised mCherry, it seems that there are also not significant differences between 1:1 and 2:1 ratios (Fig. \ref{fig:all} \textbf{a} pink and purple colours). This is surprising as, according to the other two analysed conditions, more cationised mCherry is covering the prexisting gel at 2:1 ratio, and thus the sizes of cationised mCherry regions should have been reduced at this condition too. We infer that the coverage of the gel by cationised mCherry is more extended but thinner. An equally extended but thinner cationised mCherry layer might also explain why there are no significant differences between the 5:1 and 10:1 ratios (Fig. \ref{fig:all} \textbf{a}, blue and turquoise columns). We believe this may be due to competition between cationised mCherry directly deposition on top of the pre-existing eGFP gel, and the formation of cationised mCherry clusters before their deposition on the primary gel, as illustrated in Fig. \ref{fig:all} \textbf{f}.

Lastly, we measured the sizes of only the native eGFP and cationised mCherry domains to see if their distribution also changed with the different 
ratios \paddyspeaks{of protein concentration} tested. To do this analysis, the neighbours of pixels already identified as a domain type were counted and the number of pixels per domain was obtained. 
The histograms with the distribution results are shown in Fig. \ref{fig:all} \textbf{d} and \textbf{e}, where Fig. \ref{fig:all} \textbf{d} shows the size distribution of native eGFP domains, and Fig. \ref{fig:all} \textbf{e} shows the correspondent histograms for cationised mCherry. The plots show that the amount of small areas is higher for cationised mCherry than for native eGFP, and that more areas with sizes above $\sim$ 8 $\mu$m are observed for latter. For this protein, we observe that the size distribution is smaller as the amount of cationised mCherry is decreased. This decrease in both number and size of native eGFP domains is more evident for the 5:1 and 10:1 ratios (blue and turquoise colours, respectively) as compared to equal amounts (pink) of protein and the 2:1 ratio (purple) (Fig. \ref{fig:all} \textbf{d}). This coincides with the suggestion that we obtain a larger gel coverage as the amount of cationised protein added is reduced. However, this trend is not clearly observed for cationised mCherry domains. It only seems that there is a smaller size distribution when the concentration of this protein is reduced 10-fold (turquoise), since the distribution for the rest of the concentrations looks similar, as shown in Fig. \ref{fig:all} \textbf{e}. Thus, although the actual coverage of the pre-existing gel might be manipulable, the actual size of the individual protein domains of the second protein does not seem to be so with our current proposal.

\section*{Discussion}

By 
tuning interactions in a two-component model system of proteins, we have obtained a considerable degree of control over their self-assembly. In particular, using the fluorescent proteins eGFP and mCherry as the model system allowed the identification of the conditions that preserved or affected their native structure, which is essential for future applications of 
biomaterials \paddyspeaks{based on this principle.}

When using the trivalent salt yttrium chloride, which exhibits high protein-ion specificity, percolating gels were obtained for native eGFP at \paddyspeaks{yttrium chloride} concentrations as low as 2 mM. However, for native mCherry only clusters were identified and the protein seemed to undergo reentrant solution at yttrium chloride concentrations above 5 mM, consistent with previous observations \cite{zhang2010,zhang2011,roosen-runge2013}, where the gelation of proteins with trivalent salts only occurred within a window of concentrations. Outside that range, the proteins were in solution. The studies suggested these could be due to a saturation of metallic cations in the surface of the protein, which leads to a charge inversion, making the proteins stable again and thus they re-entrant solution \cite{zhang2010,zhang2011,roosen-runge2013}. A similar saturation effect could take place with mCherry. However, this outcome is surprising 
both proteins present the same amount of acidic residues that interact with the trivalent ion \paddyspeaks{Supplementary Table S1}. A possible factor for the difference in the behaviour could be the number and surface distribution of the carboxylic groups, which might facilitate faster surface saturation or prevent proper bridging with the metal ions \cite{zhang2011}. Further studies on the charge distribution of mCherry and their spatial interactions with yttrium cations are necessary for a better understanding.

As a result, we devised an experimental approach where cationisation of mCherry was conducted to prevent its gelation with yttrium chloride and hence gain more control over the protein-salt interactions. This reaction successfully modified the net charge of the protein from negative to positive and the gelation of the cationised protein with yttrium chloride was effectively avoided. However, the non-specific interactions with ammonium sulphate were not affected and cationised mCherry gelled readily with this salt.

The desired bi-networks were obtained \paddyspeaks{by a two-step protein addition.} Firstly, a primary gel of eGFP was formed through the addition of yttrium chloride. Secondly, ammonium sulphate was added followed by cationised mCherry, which formed a gel on top of the pre-existing native eGFP structure. This coverage was not homogenous and thus individual protein domains of both native eGFP and cationised mCherry were clearly identified throughout the structure. Furthermore, coverage of the primary gel formed by native eGFP was easily manipulated by decreasing the amount of cationised mCherry added.

\ioatzinspeaks{Choosing salts with specific protein interactions and the modification of the surface of the proteins carried out above were essential for the successful decorated network formation, as} only mixed gel networks were obtained when the proteins were gelled together with different concentrations of ammonium sulphate, and no individual regions of each species were clearly distinguished. This could be due to non-specific interactions between the proteins and the salt, as well as the intrinsic similarities between eGFP and mCherry. Indeed, the isoelectric point and acidic surface residues of the two proteins are similar and only a few salt (Ca$^{+2}$ and SO$_{4}^{-2}$) and polymer (PEG) specific sites are present in eGFP, as shown in Supplementary Table II . One of these is a site for a sulphate group, which might explain why this protein gels before mCherry. Such protein similarity suggests that the interactions of both proteins with ammonium sulphate are comparable and thus they will gel at similar salt concentrations.

Our results emphasise the large complexity of protein-salt interactions and highlight the need for further studies to understand better the specific interactions involved to obtain the desired structures. In spite of this, to our knowledge, our structures are the first such decorated networks formed by two functional proteins in their native state. Binary mixtures of proteins with enzymatic, charge-carrying, antibiotic and light-harvesting abilities can be designed \emph{de novo} and finely assembled into micron-sized porous networks through the methodology developed herein. Moreover, the proteins involved in our decorated networks do not require compatibility or specificity on their mutual interactions. This is not the case for example, of immobilised enzymes on peptide hydrogels, where the proteins are fixed in the gel structure through protein-peptide interactions \cite{campbell2018}, which in turn limits the nature of both the proteins and/or peptides to be utilised. 
\paddyspeaks{Finally} our methodology is easy, cheap, versatile and scalable, and hence, it opens the door for new strategies to produce a novel class of innovative functional biomaterials.

\section*{Conflicts of interest}
There are no conflicts to declare.

\section*{Acknowledgements}
The authors would like to thank Nicholas Wood for providing the code for imaging analysis. The authors are grateful for very enriching discussions with Peter Schurtenberger and Bob Evans. IRdA was supported by a doctoral scholarship from CONACyT. \paddyspeaks{IRdA, AC-P and CPR gratefully acknowldge the ERC Grant agreement n$^\circ$ 617266 ``NANOPRS'' for financial support and \paddyspeaks{Engineering and Physical Sciences Research Council (EP/H022333/1).}} 

\subsection*{Methods} 

Fluorescent proteins were obtained through bacterial expression, followed by cell separation, cell lysis, centrifugation, selective purification of the proteins and concentration. Further protein cationisation was performed to change the negative charge of the native proteins to a positive one without affecting the protein functionality. Protein gelation was carried out by adding different amounts of ammonium sulphate, iron chloride and yttrium chloride. Finally, image analysis was performed to quantify the size of individual protein domains within the decorated gel.

\subsection{Cellular Culture for the Expression of eGFP}
\label{subsec:Exp-GFP}
\emph{Escherichia coli} BL21 (DE3) competent (able to receive DNA) cells were previously transformed with the DNA plasmid-pET45b(+)-eGFP. First, a mini-culture was prepared by inoculating 100 mL of lysogeny broth (LB) of nutrients and the antibiotic carbenicillin (50 $\mu$g/mL) with \emph{E. coli.}
The culture was left to grow for 16 h at 37$^{\circ}$ C and 180 rpm. 20 mL of this culture were then used to inoculate 1 L of LB containing carbenicillin (50 $\mu$g/mL), which was left to grow under the same previous conditions. The optical density (OD$_{600nm}$) was monitored to a value of 0.5-0.6, when the production of eGFP was induced by adding 1 mM of Isopropyl $\beta$-D-1-thiogalactopyranoside (IPTG) \cite{beckwith1967}. After 1 h of induction time, the temperature was changed to 30$^{\circ}$ C. After 16 h of incubation, the cell culture was centrifuged  at 4500 g for 15 min at 4$^{\circ}$ C. The supernantant obtained was discarded and the pellet was resuspended in a \emph{lysis buffer} (20 mM imidazole, 300 mM NaCl and 50 mM potassium phosphate at pH 8.0) and stored at -20$^{\circ}$ C.

\subsection{Cellular Culture for the Expression of mCherry}
\label{subsec:Exp-mCherry}

\emph{Escherichia coli} BL21 (DE3) competent cells were previously transformed with the DNA plasmid pBADmCherry. A mini-culture was prepared under the same conditions as for the expression of eGFP. Equally, 20 mL of said culture were used to inoculate 1 L of LB with 50 $\mu$g/mL carbenicillin, which was then incubated as before. However, in this case the OD$_\mathrm{600nm}$ of the culture was closely monitored until it reached a value between 0.6-0.8, when mCherry expression was induced by adding arabinose from a 20\% stock solution for a final concentration of 0.2\% \cite{schleif2000}. The rest of the steps followed are the same as the ones described above for eGFP.

\subsection{Purification of Fluorescent Proteins}
\label{subsec:Purification-proteins}

The purification of both proteins followed the same protocol. Cell pellets were \paddyspeaks{\textbf{defrosted}} \paddycomment{check with Ross!} and kept on ice, broken down by lysis using 3 sonication cycles of 30 seconds (Soniprep 150 plus MSE) and centrifuged at 18000 rpm (Sorvall SS34 rotor) at 4$^{\circ}$ C for 30 min. The supernatant was recovered and filtered through a 0.22 $\mu$m syringe filter (Millipore) and injected to a Ni-NTA agarose column (Qiagen) connected to an \"AKTA start purification system (GE Healthcare). The column had been previously equilibrated with the lysis buffer mentioned above in section \ref{subsec:Exp-GFP}. After the protein solution addition, the bound fluorescent proteins were washed with the same lysis buffer to elute unbound proteins. eGFP and mCherry were later eluted with a linear gradient (0-100\%) of a 500 mM imidazole, 300 mM NaCl, 50 mM potassium phosphate buffer at pH 8. The recovered proteins were dialysed against dH$_{2}$O for 16 h using a 10MWCo dialysis membrane. Finally, the proteins were collected and stored at -20$^{\circ}$ C.

\subsection{Concentration of Fluorescent Proteins}
\label{subsec:Concentration-proteins}

Dialysed proteins were filtered through 0.22 $\mu$m syringe filter (Millipore) and concentrated by reducing a volume of \ioatzinspeaks{$\sim$ 30 ml} to $\sim$1 ml using protein 30 kDa concentrators (ThermoFisher Scientific) at 5000 rpm and 4$^{\circ}$ C for the time required to reach the desired volume. The protein concentration was determined by measuring the absorbance at $\lambda_\mathrm{eGFP}$ = 488 nm and $\lambda_\mathrm{mCherry}$ = 587 nm, using molar extinction coefficients values of $\epsilon_\mathrm{eGFP}$ = 56000 M$^{-1}$cm$^{-1}$ \cite{kaishima2016} and  $\epsilon_\mathrm{mCherry}$ = 72000 M$^{-1}$cm$^{-1}$ \cite{li2007}.

\subsection{Gelation of Native and Cationised eGFP and mCherry with Yttrium Chloride}
\label{subsec:Fluo-YCl3}

Different yttrium chloride (YCl$_3$) concentrations (1, 2, 5, 10, 50 mM) were added from stock solutions to 30 $\mu$L of 7 mg/mL solutions of native and cationised proteins, separately. The solutions were mixed in a vortex for 5 min and analysed immediately after.

\subsection{Gelation of eGFP and Native and Cationised mCherry with Ammonium Sulphate}
\label{subsec:PP-prot-NH42SO4}

Different amounts of ammonium sulphate, (NH$_\mathrm{4}$)$_\mathrm{2}$SO$_\mathrm{4}$, Sigma Aldrich) were increasingly added to a 100 $\mu$L of 14 mg/mL total mixture of native eGFP and native mCherry to reach final concentrations from 0.3 to 3 M.  Additionally, to test its effect on modified proteins, (NH$_\mathrm{4}$)$_\mathrm{2}$SO$_\mathrm{4}$ was added to 50 $\mu$L of 7 mg/mL solutions of cationised mCherry (see section below) for a final concentration of 3 M of the salt. All samples were vortexed for 5 min and analysed immediately after.

\subsection{Cationisation of mCherry}
\label{subsec:Prot-Cat}

Protein cationisation was performed through the addition of  a stock solution of hexamethylendiamine (HMDA, Sigma Aldrich) at a pH 6.0-6.5 whose concentration was 10 times that of the protein solution to 10 mL of a known concentration of native protein solution. The pH was adjusted to 6.0-6.5 with HCl 1 M. An equal concentration to the protein of 1-Ethyl-3-(3-dimethylaminopropyl)carbodiimide (EDC, Sigma Aldrich) was added to the reaction at two different times: half of the required reagent was added after HMDA and the remaining half after $\sim$3h of reaction. The pH was monitored constantly and adjusted to 6.0-6.5 as required for the first 6 h. The mixture was left stirring at room temperature overnight ($\sim$18 h). Finally, the solution was filtered through a 0.22 $\mu$m syringe filter to remove any precipitates, dialysed against dH$_\mathrm{2}$O for 24 h and concentrated following the procedure in section \ref{subsec:Concentration-proteins}. The degree of cationisation was determined using zeta potential measurements (Zetasizer Nano ZS, Malvern) using 1 mL of 2 mg/mL of native and modified proteins.

\subsection{Decorated Gel Network Formation of Native and Cationised Proteins with Yttrium Chloride and Ammonium Sulphate}
\label{subsec:Fluo-prot}

Firstly, an eGFP gel was formed by adding 5mM YCl$_\mathrm{3}$ as described above to a 5 mg/mL solution of only eGFP. Then (NH$_\mathrm{4}$)$_\mathrm{2}$SO$_\mathrm{4}$ was added and dissolved for a final concentration of 3 M. Finally, 5 mg/mL of cationised mCherry were added to the solution, mixed for 5 min and analysed immediately. The final total protein concentration was kept to 10 mg/mL. Three more samples following this protocol were prepared where the mass ratio of native eGFP:cationised mCherry was varied to 2:1, 5:1 and 10:1. 



\subsection{Imaging of the protein gels}
\label{subsec:Charac-Prot}

All samples were confined to capillaries with a square cross-section of of 0.50 x 0.50 mm (Vitocrom) and sealed on the ends with Norland61. Confocal laser scanning microscopy Leica TCS with a white light laser emitting at 500 nm was used to study any gelation, using a NA 1.4 63x oil immersion objective. The channels used for the proteins were 488 nm for eGFP and 587 nm for native and cationised mCherry. Scans of the capillary in the $z$-axis were also acquired to analyse the gel structures in 3d, where care was taken to assure the pixel size was equal in all axes (100 nm/pixel).

\subsection{Analysis of the Mixing and Domain Sizes of Native eGFP and Cationised mCherry in the Gel Bi-networks}
\label{subsec:Analysis-Prot}

Confocal optical slices of the obtained gels were analysed individually to measure the percentage of mixing and de-mixing of the gels, along with the sizes of de-mixed domains. Care was taken to obtain images with the same sizes and the intensities of the images was normalised for all of them. Then, the pixels were classified according to their intensities as yellow (mixed proteins), green (eGFP) and red (cationised mCherry). Different intensity thresholds were tested to optimise the results. The percentage of each colour was obtained to study the mixing of the proteins. Additionally, the areas of the identified regions of individual proteins (de-mixed domains) were measured by counting the number of pixels on said domains, which were then converted to $\mu$m$^{2}$ using the pixel size. Regions with sizes below 10 pixels, which corresponded to 1\% of the total size of the image, were discarded.




\begin{thebibliography}{10}
\expandafter\ifx\csname url\endcsname\relax
  \def\url#1{\texttt{#1}}\fi
\expandafter\ifx\csname urlprefix\endcsname\relax\def\urlprefix{URL }\fi
\providecommand{\bibinfo}[2]{#2}
\providecommand{\eprint}[2][]{\url{#2}}

\bibitem{xu2016}
\bibinfo{author}{Zongwei, X.}, \bibinfo{author}{Liyang, W.},
  \bibinfo{author}{Fengzhou, F.}, \bibinfo{author}{Yongqi, F.} \&
  \bibinfo{author}{Yin, Z.}
\newblock \bibinfo{title}{A review on colloidal self-assembly and their
  applications}.
\newblock \emph{\bibinfo{journal}{Current Nanoscience}}
  \textbf{\bibinfo{volume}{12}}, \bibinfo{pages}{725} (\bibinfo{year}{2016}).

\bibitem{bai2016}
\bibinfo{author}{Bai, Y.}, \bibinfo{author}{Luo, Q.} \& \bibinfo{author}{Liu,
  J.}
\newblock \bibinfo{title}{Protein self-assembly via supramolecular strategies}.
\newblock \emph{\bibinfo{journal}{Chemical Society Reviews}}
  \textbf{\bibinfo{volume}{45}}, \bibinfo{pages}{2756} (\bibinfo{year}{2016}).

\bibitem{luo2016}
\bibinfo{author}{Luo, Q.}, \bibinfo{author}{Hou, C.}, \bibinfo{author}{Bai,
  Y.}, \bibinfo{author}{Wang, R.} \& \bibinfo{author}{Liu, J.}
\newblock \bibinfo{title}{Protein assembly: Versatile approaches to construct
  highly ordered nanostructures}.
\newblock \emph{\bibinfo{journal}{Chemical Reviews}}
  \textbf{\bibinfo{volume}{116}}, \bibinfo{pages}{13571}
  (\bibinfo{year}{2016}).

\bibitem{kobayashi2018}
\bibinfo{author}{Kobayashi, N.} \emph{et~al.}
\newblock \bibinfo{title}{Self-assembling supramolecular nanostructures
  constructed from de novo extender protein nanobuilding blocks}.
\newblock \emph{\bibinfo{journal}{ACS Synthetic Biology}}
  \textbf{\bibinfo{volume}{7}}, \bibinfo{pages}{1381} (\bibinfo{year}{2018}).

\bibitem{yang2016}
\bibinfo{author}{Yang, L.} \emph{et~al.}
\newblock \bibinfo{title}{Self-assembly of proteins: Towards supramolecular
  materials}.
\newblock \emph{\bibinfo{journal}{Chemistry --A European Journal}}
  \textbf{\bibinfo{volume}{22}}, \bibinfo{pages}{15570} (\bibinfo{year}{2016}).

\bibitem{vo-dinh2005}
\bibinfo{author}{Vo-Dinh, T.}
\newblock \emph{\bibinfo{title}{Protein Nanotechnology}},
  \bibinfo{pages}{1--13} (\bibinfo{publisher}{Humana Press},
  \bibinfo{address}{Totowa, NJ}, \bibinfo{year}{2005}).

\bibitem{watkins2017}
\bibinfo{author}{Watkins, D.~W.} \emph{et~al.}
\newblock \bibinfo{title}{Construction and in vivo assembly of a catalytically
  proficient and hyperthermostable de novo enzyme}.
\newblock \emph{\bibinfo{journal}{Nature Communications}}
  \textbf{\bibinfo{volume}{8}}, \bibinfo{pages}{358} (\bibinfo{year}{2017}).

\bibitem{dumetz2008}
\bibinfo{author}{Dumetz, A.}, \bibinfo{author}{Chockla, A.~M.},
  \bibinfo{author}{Kaler, E.~W.} \& \bibinfo{author}{Lenhoff, A.~M.}
\newblock \bibinfo{title}{Protein phase behavior in aqueous solutions:
  Crystallization, liquid-liquid phase separation, gels, and aggregates}.
\newblock \emph{\bibinfo{journal}{Biophysical Journal}}
  \textbf{\bibinfo{volume}{94}}, \bibinfo{pages}{570} (\bibinfo{year}{2008}).

\bibitem{mcmanus2016}
\bibinfo{author}{McManus, J.~J.}, \bibinfo{author}{Charbonneau, P.},
  \bibinfo{author}{Zaccarelli, E.} \& \bibinfo{author}{Asherie, N.}
\newblock \bibinfo{title}{The physics of protein self-assembly}.
\newblock \emph{\bibinfo{journal}{Current Opinion in Colloid \& Interface
  Science}} \textbf{\bibinfo{volume}{22}}, \bibinfo{pages}{73}
  (\bibinfo{year}{2016}).

\bibitem{zhang2007}
\bibinfo{author}{Zhang, F.} \emph{et~al.}
\newblock \bibinfo{title}{Protein interactions studied by saxs: Effect of ionic
  strength and protein concentration for bsa in aqueous solutions}.
\newblock \emph{\bibinfo{journal}{The Journal of Physical Chemistry B}}
  \textbf{\bibinfo{volume}{111}}, \bibinfo{pages}{251} (\bibinfo{year}{2007}).

\bibitem{wang2012}
\bibinfo{author}{Wang, Y.} \emph{et~al.}
\newblock \bibinfo{title}{Pathological crystallization of human
  immunoglobulins}.
\newblock \emph{\bibinfo{journal}{Proceedings of the National Academy of
  Sciences}} \textbf{\bibinfo{volume}{109}}, \bibinfo{pages}{13359}
  (\bibinfo{year}{2012}).

\bibitem{saric2014}
\bibinfo{author}{{\v S}ari{\'c}, A.}, \bibinfo{author}{Chebaro, Y.~C.},
  \bibinfo{author}{Knowles, T. P.~J.} \& \bibinfo{author}{Frenkel, D.}
\newblock \bibinfo{title}{Crucial role of nonspecific interactions in amyloid
  nucleation}.
\newblock \emph{\bibinfo{journal}{Proceedings of the National Academy of
  Sciences}} \textbf{\bibinfo{volume}{111}}, \bibinfo{pages}{17869}
  (\bibinfo{year}{2014}).

\bibitem{zaccarelli2007}
\bibinfo{author}{Zaccarelli, E.}
\newblock \bibinfo{title}{Colloidal gels: equilibrium and non-equilibrium
  routes}.
\newblock \emph{\bibinfo{journal}{Journal of Physics: Condensed Matter}}
  \textbf{\bibinfo{volume}{19}}, \bibinfo{pages}{323101}
  (\bibinfo{year}{2007}).

\bibitem{varrato2012}
\bibinfo{author}{Varrato, F.} \emph{et~al.}
\newblock \bibinfo{title}{Arrested demixing opens route to bigels}.
\newblock \emph{\bibinfo{journal}{Proceedings of the National Academy of
  Sciences}} \textbf{\bibinfo{volume}{109}}, \bibinfo{pages}{19155}
  (\bibinfo{year}{2012}).

\bibitem{royall2018}
\bibinfo{author}{Royall, C.~P.}, \bibinfo{author}{Williams, S.~R.} \&
  \bibinfo{author}{Tanaka, H.}
\newblock \bibinfo{title}{Vitrification and gelation in sticky spheres}.
\newblock \emph{\bibinfo{journal}{The Journal of Chemical Physics}}
  \textbf{\bibinfo{volume}{148}}, \bibinfo{pages}{044501}
  (\bibinfo{year}{2018}).

\bibitem{whitelam2014}
\bibinfo{author}{Whitelam, S.}, \bibinfo{author}{Hedges, L.~O.} \&
  \bibinfo{author}{Schmit, J.~D.}
\newblock \bibinfo{title}{Self-assembly at a nonequilibrium critical point}.
\newblock \emph{\bibinfo{journal}{Physical Review Letters}}
  \textbf{\bibinfo{volume}{112}}, \bibinfo{pages}{155504}
  (\bibinfo{year}{2014}).

\bibitem{dimichele2013}
\bibinfo{author}{Di~Michele, L.} \emph{et~al.}
\newblock \bibinfo{title}{Multistep kinetic self-assembly of dna-coated
  colloids}.
\newblock \emph{\bibinfo{journal}{Nature Communications}}
  \textbf{\bibinfo{volume}{4}}, \bibinfo{pages}{2007} (\bibinfo{year}{2013}).

\bibitem{blumlein2015}
\bibinfo{author}{Blumlein, A.} \& \bibinfo{author}{McManus, J.~J.}
\newblock \bibinfo{title}{Bigels formed via spinodal decomposition of unfolded
  protein}.
\newblock \emph{\bibinfo{journal}{Journal of Materials Chemistry B}}
  \textbf{\bibinfo{volume}{3}}, \bibinfo{pages}{3429} (\bibinfo{year}{2015}).

\bibitem{elcock2001}
\bibinfo{author}{Elcock, A.~H.} \& \bibinfo{author}{McCammon, J.~A.}
\newblock \bibinfo{title}{Calculation of weak protein-protein interactions: The
  ph dependence of the second virial coefficient}.
\newblock \emph{\bibinfo{journal}{Biophysical Journal}}
  \textbf{\bibinfo{volume}{80}}, \bibinfo{pages}{613} (\bibinfo{year}{2001}).

\bibitem{duong-ly2014}
\bibinfo{author}{Duong-Ly, K.~C.}, \bibinfo{author}{Gabelli, S.~B.} \&
  \bibinfo{author}{Lorsch, J.}
\newblock \emph{\bibinfo{title}{Chapter Seven - Salting out of Proteins Using
  Ammonium Sulfate Precipitation}}, vol. \bibinfo{volume}{541},
  \bibinfo{pages}{85} (\bibinfo{publisher}{Academic Press},
  \bibinfo{year}{2014}).

\bibitem{okur2017}
\bibinfo{author}{Okur, H.~I.} \emph{et~al.}
\newblock \bibinfo{title}{Beyond the hofmeister series: Ion-specific effects on
  proteins and their biological functions}.
\newblock \emph{\bibinfo{journal}{The Journal of Physical Chemistry B}}
  \textbf{\bibinfo{volume}{121}}, \bibinfo{pages}{1997} (\bibinfo{year}{2017}).

\bibitem{piazza2000}
\bibinfo{author}{Piazza, R.} \& \bibinfo{author}{Pierno, M.}
\newblock \bibinfo{title}{Protein interactions near crystallization: a
  microscopic approach to the hofmeister series}.
\newblock \emph{\bibinfo{journal}{Journal of Physics: Condensed Matter}}
  \textbf{\bibinfo{volume}{12}}, \bibinfo{pages}{A443--A449}
  (\bibinfo{year}{2000}).
\newblock \urlprefix\url{https://doi.org/10.1088%2F0953-8984%2F12%2F8a%2F361}.

\bibitem{zhang2010}
\bibinfo{author}{Zhang, F.} \emph{et~al.}
\newblock \bibinfo{title}{Universality of protein reentrant condensation in
  solution induced by multivalent metal ions}.
\newblock \emph{\bibinfo{journal}{Proteins: Structure, Function, and
  Bioinformatics}} \textbf{\bibinfo{volume}{78}}, \bibinfo{pages}{3450}
  (\bibinfo{year}{2010}).

\bibitem{zhang2011}
\bibinfo{author}{Zhang, F.}, \bibinfo{author}{Zocher, G.},
  \bibinfo{author}{Sauter, A.}, \bibinfo{author}{Stehle, T.} \&
  \bibinfo{author}{Schreiber, F.}
\newblock \bibinfo{title}{Novel approach to controlled protein crystallization
  through ligandation of yttrium cations}.
\newblock \emph{\bibinfo{journal}{Journal of Applied Crystallography}}
  \textbf{\bibinfo{volume}{44}}, \bibinfo{pages}{755} (\bibinfo{year}{2011}).

\bibitem{roosen-runge2013}
\bibinfo{author}{Roosen-Runge, F.}, \bibinfo{author}{Heck, B.~S.},
  \bibinfo{author}{Zhang, F.}, \bibinfo{author}{Kohlbacher, O.} \&
  \bibinfo{author}{Schreiber, F.}
\newblock \bibinfo{title}{Interplay of ph and binding of multivalent metal
  ions: Charge inversion and reentrant condensation in protein solutions}.
\newblock \emph{\bibinfo{journal}{The Journal of Physical Chemistry B}}
  \textbf{\bibinfo{volume}{117}}, \bibinfo{pages}{5777} (\bibinfo{year}{2013}).

\bibitem{zhang2014b}
\bibinfo{author}{Zhang, F.} \emph{et~al.}
\newblock \bibinfo{title}{Reentrant condensation, liquid-liquid phase
  separation and crystallization in protein solutions induced by multivalent
  metal ions}.
\newblock \emph{\bibinfo{journal}{Pure and Applied Chemistry}}
  \textbf{\bibinfo{volume}{86}}, \bibinfo{pages}{191} (\bibinfo{year}{2014}).

\bibitem{wingfield2001}
\bibinfo{author}{Wingfield, P.~T.}
\newblock \bibinfo{title}{Protein precipitation using ammonium sulfate}.
\newblock \emph{\bibinfo{journal}{Current Protocols in Protein Science}}
  \bibinfo{pages}{Appendix--3F} (\bibinfo{year}{2001}).

\bibitem{konno2001}
\bibinfo{author}{Konno, T.}
\newblock \bibinfo{title}{Amyloid-induced aggregation and precipitation of
  soluble proteins: An electrostatic contribution of the alzheimer's
  beta(25−35) amyloid fibril}.
\newblock \emph{\bibinfo{journal}{Biochemistry}} \textbf{\bibinfo{volume}{40}},
  \bibinfo{pages}{2148} (\bibinfo{year}{2001}).

\bibitem{campbell2018}
\bibinfo{author}{Campbell, E.~C.} \emph{et~al.}
\newblock \bibinfo{title}{Hydrogel-immobilized supercharged proteins}.
\newblock \emph{\bibinfo{journal}{Advanced Biosystems}}
  \textbf{\bibinfo{volume}{2}}, \bibinfo{pages}{1700240}
  (\bibinfo{year}{2018}).

\bibitem{beckwith1967}
\bibinfo{author}{Beckwith, J.~R.}
\newblock \bibinfo{title}{Regulation of the lac operon}.
\newblock \emph{\bibinfo{journal}{Science}} \textbf{\bibinfo{volume}{156}},
  \bibinfo{pages}{597} (\bibinfo{year}{1967}).

\bibitem{schleif2000}
\bibinfo{author}{Schleif, R.}
\newblock \bibinfo{title}{Regulation of the l-arabinose operon of escherichia
  coli}.
\newblock \emph{\bibinfo{journal}{Trends in Genetics}}
  \textbf{\bibinfo{volume}{16}}, \bibinfo{pages}{559} (\bibinfo{year}{2000}).

\bibitem{kaishima2016}
\bibinfo{author}{Kaishima, M.}, \bibinfo{author}{Ishii, J.},
  \bibinfo{author}{Matsuno, T.}, \bibinfo{author}{Fukuda, N.} \&
  \bibinfo{author}{Kondo, A.}
\newblock \bibinfo{title}{Expression of varied gfps in saccharomyces
  cerevisiae: codon optimization yields stronger than expected expression and
  fluorescence intensity}.
\newblock \emph{\bibinfo{journal}{Scientific Reports}}
  \textbf{\bibinfo{volume}{6}}, \bibinfo{pages}{35932} (\bibinfo{year}{2016}).

\bibitem{li2007}
\bibinfo{author}{Li, Y.}, \bibinfo{author}{Sierra, A.~M.}, \bibinfo{author}{Ai,
  H.-w.} \& \bibinfo{author}{Campbell, R.~E.}
\newblock \bibinfo{title}{Identification of sites within a monomeric red
  fluorescent protein that tolerate peptide insertion and testing of
  corresponding circular permutations}.
\newblock \emph{\bibinfo{journal}{Photochemistry and Photobiology}}
  \textbf{\bibinfo{volume}{84}}, \bibinfo{pages}{111} (\bibinfo{year}{2007}).

\end{thebibliography}

\end{document}